\documentclass[%
 reprint,
superscriptaddress,
nofootinbib,
 amsmath,amssymb,floatfix,
]{revtex4-1}

\usepackage{bm}
\usepackage{graphicx, color}
\usepackage[export]{adjustbox}
\usepackage{gensymb}

\newcommand{\Pud}{P^\mathrm{ud}}
\newcommand{\Pglobal}{P^\mathrm{global}}
\newcommand{\rhocoex}{\rho_\mathrm{coex}}
\newcommand{\Pcoex}{P_\mathrm{coex}}

\newcommand{\partdev}[2]{\frac{\partial #1}{\partial #2}}
\newcommand{\diffix}[3]{\left(\frac{\partial #1}{\partial #2}\right)_{#3}}

\begin{document}

\title{A simple and accurate method to determine  fluid-crystal phase boundaries from direct coexistence simulations}


\author{Frank Smallenburg}
\email{frank.smallenburg@cnrs.fr}
\affiliation{Universit\'e Paris-Saclay, CNRS, Laboratoire de Physique des Solides, 91405 Orsay, France}
\author{Giovanni Del Monte}
\affiliation{Soft Condensed Matter and Biophysics, Debye Institute of Nanomaterials Science, Utrecht University, Utrecht, Netherlands}
\author{Marjolein de Jager}
\affiliation{Soft Condensed Matter and Biophysics, Debye Institute of Nanomaterials Science, Utrecht University, Utrecht, Netherlands}
\author{Laura Filion}
\affiliation{Soft Condensed Matter and Biophysics, Debye Institute of Nanomaterials Science, Utrecht University, Utrecht, Netherlands}

\date{\today}

\begin{abstract}
One method for computationally determining phase boundaries is to explicitly simulate a direct coexistence between the two phases of interest. Although this approach works very well for fluid-fluid coexistences, it is often considered to be less useful for fluid-crystal transitions, as additional care must be taken to prevent the simulation boundaries from imposing unwanted strains on the crystal phase. Here, we present a simple adaptation to the direct coexistence method that nonetheless allows us to obtain highly accurate predictions of fluid-crystal coexistence conditions, assuming a fluid-crystal interface can be readily simulated. We test our approach on hard spheres, the screened Coulomb potential, and a 2D patchy-particle model. In all cases, we find excellent agreement between the direct coexistence approach and (much more cumbersome) free-energy calculation methods. Moreover, the method is sufficiently accurate to resolve the (tiny) free-energy difference between the face-centered cubic and hexagonally close-packed crystal of hard spheres in the thermodynamic limit.  The simplicity of this method also ensures that it can be trivially implemented in essentially any simulation method or package. Hence, this approach provides an excellent alternative to free-energy based methods for the precise determination of phase boundaries.
\end{abstract}

\maketitle 

\section{Introduction}

Phase transitions between a disordered fluid phase and an ordered crystal are of paramount importance to a wide range of physical phenomena, including colloidal self-assembly, ice formation in water, and the melting, solidification, and interfacial behavior of a vast array of molecular and atomic substances. When studying these phenomena in computer simulations, a key first step is inevitably the determination of the phase boundary: under what conditions can the fluid and crystal phase coexist, i.e. have the same temperature, pressure, and chemical potential?

A large number of methods have been introduced that use computer simulations to address this question \cite{bookfrenkel,chew2023phase}. Although exceptions exist (e.g. \cite{bruce2000lattice, wilding2000freezing, chen2001direct}), these methods can broadly  be grouped in three different categories. The first category is to simply explore which phase emerges from a simulation performed at a specific state point. Since fluid-crystal transitions are nearly always first-order phase transitions, the effectiveness of this method is typically hindered by hysteresis: fluids can be supercooled and solids superheated. As a result, spontaneous phase transitions are rarely observed at the equilibrium melting or freezing point. Nonetheless, this approach can be extremely useful to obtain a rough impression of the phase behavior of a new system.

The second category consists of free-energy based methods, typically involving some form of thermodynamic integration \cite{frenkel1984new,vega2008determination,bookfrenkel}. 
In many cases, this involves determining the free energy of each phase and then finding the state points where the temperatures, pressures, and chemical potentials of the two phases are equal. Calculating the free energy of a fluid is typically straightforward, and can be done via thermodynamic integration over the equation of state, using the ideal gas as a reference system \cite{bookfrenkel}. For the crystal phase, more advanced methods are needed, involving more complex integration pathways. Arguably the most standard approach is an integration from the Einstein crystal introduced by Frenkel and Ladd \cite{frenkel1984new}. A large number of variations and extensions to this approach have been developed, both attempting to optimize the method and to extend it to different systems and phases (see e.g. \cite{bolhuis1997tracing, schilling2009computing, polson2000finite, vega2007revisiting, dijkstra2013phase, moir2021tethered, vega2008determination}). The advantage of this class of methods is that generally each individual simulation only samples a single phase, avoiding the need for explicit interfaces. Historically, this has been an important benefit as it allows obtaining accurate results from relatively small simulation sizes with short simulation times. As a downside, this approach requires integration over a (or usually multiple) series of simulation results, where the results can be influenced by e.g. the number of state points sampled and the chosen integration limits. As a result, the barrier to actually performing a full free-energy calculation for a given system is significant, and hence their application is usually limited to fundamental models where the effort is deemed warranted. 

The third category are direct coexistence simulations.  Dating back to the 1970s \cite{opitz1974molecular, ladd1977triple, ladd1978interfacial, cape1978molecular}, these are simulations which incorporate an explicit interface between the fluid and solid. In principle, the exchange of particles, volume, and energy between the two phases then directly imposes the conditions for coexistence. However, in the case of a fluid-crystal system, this approach is complicated by the fact that a crystal can sustain a strain, and is therefore sensitive to the shape and size of the simulation box that confines it \cite{broughton1986molecular}. Clearly the equilibrium crystal should be unstrained, and multiple methods have been developed to ensure a strain-free crystal. The first attempts to do this simply required that the overall pressure tensor in the direct coexistence simulation was isotropic, an approach that has been applied in a variety of ensembles (see e.g. \cite{morris2002melting, morris1994melting, yoo2004melting, lanning2004solid, wang2005melting, garcia2006melting}). Technically, this is not correct, since the presence of an interface also provides an anisotropic contribution to the overall pressure tensor. Instead, the goal should be to ensure that the pressure tensor inside the crystal phase is isotropic. One method to address this in the microcanonical ($NVE$) or canonical ($NVT$) ensemble is to measure the local pressure tensor inside the coexisting crystal phase and adjusting the simulation box to ensure that it is isotropic \cite{davidchack1998simulation}. Another, more commonly used, approach is to perform simulations in a thermodynamic ensemble  where number of particles $N$ and temperature $T$ are fixed, and the size of the simulation box is only allowed to fluctuate in the direction perpendicular to the interfaces, controlled by a pressure $P_z$ \cite{noya2008determination, espinosa2013fluid, zykova2010monte}. In this $NP_zT$ ensemble, the shape of the box along the other two directions is kept fixed in accordance with the lattice parameters of the crystal at an isotropic pressure $P=P_z$. The downside of a constant-pressure ensemble is the fact that the fluid-crystal interface is no longer stable: eventually, the crystal will either melt or fully fill the simulation box. The coexistence conditions must therefore be determined by finding the pressure where the crystal has an equal probability of growing or shrinking, which may require a large number of long simulations and introduces a stochastic complication to the process. A solution was proposed by Pedersen \textit{et al.} \cite{pedersen2013computing} in the form of interface pinning simulations, where the interface is pinned in place via a biasing potential based on the degree of crystalline order in the system. In this approach, coexistence conditions are determined by finding the pressure at which the effective force exerted by the biasing potential vanishes. Although this approach avoids the stochasticity and long simulation times of the direct $NP_zT$ approach, it also adds an additional complication in the form of a biasing potential and the need for a suitable order parameter to determining crystallinity.

Here, we propose an elegant, accurate, and efficient method to determine fluid-crystal coexistence conditions in the $NVT$ ensemble. It relies only on global measurements of standard thermodynamic quantities, without requiring any biasing, numerical integration, or reference states. We test this method by applying it to three model systems: the hard-sphere model, a point Yukawa model, and a two-dimension patchy-particle model. In all cases, we find excellent agreement between our proposed method and either literature values or our own predictions based on thermodynamic integration. For the hard-sphere model in  particular, we show that the accuracy of our method is sufficiently high to resolve the small free-energy difference (approx. $0.001 k_B T$ per particle) between the face-centered cubic and hexagonally close-packed phases.

\section{Models}

We consider fluid-crystal coexistence in three model systems: hard spheres, Yukawa particles, and patchy particles. Here, we describe these models in detail.

\subsection{Hard spheres}

An ideal model system for testing methods to determine phase boundaries is the hard-sphere model, as the phase behavior has been extensively studied using a variety of methods (see Ref. \onlinecite{royall2023colloidal} for an overview). The hard-sphere model consists of spheres of diameter $\sigma$ which are not allowed to overlap, but otherwise have no interaction. Its phase behavior consists of a fluid at densities below the freezing density $\rhocoex^F \sigma^3 \simeq 0.939$, a face-centered cubic crystal above the melting density $\rhocoex^X \sigma^3 \simeq 1.037$, and a coexistence region in between. The corresponding coexistence pressure is $\beta P \sigma^3 \simeq 11.56$, where $\beta = 1/k_B T$, with $k_B T$ the thermal energy.

We simulate systems of $N$ hard spheres of identical mass $m$ and diameter $\sigma$ in a volume $V$, using the EDMD simulation code of Ref. \onlinecite{smallenburg2022efficient}, adapted to measure the pressure tensor. We do not make use of a thermostat, and hence the total energy of the system (which consists only of the kinetic energy) is fixed. This in turn also fixes the temperature $T$.
During the simulation, we measure the pressure tensor $P_{ij}$ by keeping track of the momentum transfer during each collision, and using the expression:
\begin{equation}\label{eq:HS-pressure}
P_{ij} = \rho k_B T \delta_{ij} - \frac{1}{V} \frac{\sum_{k} m \, \delta v_{i}^{(k)} \, \delta 
r_{j}^{(k)}}{t_\mathrm{end} - t_\mathrm{start}},
\end{equation}
where $\delta_{ij}$ is the Kronecker delta, $\rho = N/V$ the number density, and $k_B$ Boltzmann's constant. The sum runs over all collisions $k$ occurring between times $t_\mathrm{start}$ and $t_\mathrm{end}$. For each collision, $\delta 
\mathbf{r}^{(k)}$ and $ \delta \mathbf{v}^{(k)}$ denote the relative position and velocity of the two particles involved in the collision, respectively.

\subsection{Yukawa particles}

As our second model, we consider point particles interacting via the Yukawa (or screened Coulomb) potential, given by
\begin{equation}
    V_\mathrm{Yuk}(r) = \epsilon \frac{\exp(-\kappa(r-\sigma))}{r/\sigma},
\end{equation}
with $\sigma$ an effective particle size, $\epsilon$ the contact value of the potential at $r = \sigma$, and $\kappa$ the inverse screening length. In particular, we focus on a system with an inverse screening length $\kappa \sigma = 4$, and a contact value $\epsilon/k_B T = 20$, which is known to form a body-centered cubic (BCC) crystal phase upon freezing \cite{hynninen2003phase}. The interaction potential was truncated and shifted to zero at a cutoff distance $r_c = 4.5 \sigma$. 

We simulate these particles using the LAMMPS simulation package\cite{thompson2022lammps,brown2011implementing}. 
The integration time step was set to $dt=5\cdot10^{-3}\tau$. As a thermostat, we use Nos\'e-Hoover chains with 30 oscillators in the chain and a damping parameter $\tau_d = 2.0\tau$. An example script is provided in the supplemental material.

\subsection{Patchy disks}
The third model we consider is an example of an anisotropic model: a two-dimensional system of patchy particles. Depending on the number and size of the attractive patches, patchy particles in two dimensions can form a variety of (quasi)crystalline structures \cite{doye2007controlling, doppelbauer2010self, van2012formation}.
For simplicity, we focus on four-patch particles, modeled using the Kern-Frenkel potential \cite{kern2003fluid}, involving a hard core repulsion and 4 directional attractive patches, whose angular position is evenly spaced.
Specifically, the interaction potential is given by:
\begin{equation}
V_{KF}(\mathbf{r}_{ij}, \theta_i, \theta_j)=V^{\textsc{HS}}(r_{ij})+V^{\textsc{SW}}({r}_{ij})f(\mathbf{r}_{ij},\theta_i, \theta_j)),
\end{equation}
where $r_{ij} = |\mathbf{r}_{ij}|$ is the center-to-center distance between particles $i$ and $j$, and $\theta_i$ denotes the orientation of particle $i$. Additionally, $V^{\textsc{HS}}$ is the hard-disk potential with diameter $\sigma$, and $V^{\textsc{SW}}$ is a square-well potential, given by
\begin{equation}
V^{\textsc{SW}}(r)= \begin{cases} 
\epsilon & r \leq \lambda_p \\
0 & r > \lambda_p,
\end{cases}
\end{equation}
where we choose the interaction range $\lambda_p = 1.12 \sigma$ and attractive strength $\epsilon=-3k_BT$.
Finally, $f(\mathbf{r}_{ij},\theta_i, \theta_j)$ specifies the directionality of the interactions: 
\begin{equation}
f(\mathbf{r}_{ij},\theta_i, \theta_j)=\begin{cases}
1  & \begin{cases}\hat{\mathbf{n}}_{\alpha}^{(i)} \cdot \hat{\mathbf{r}}_{ij} > \cos \theta \text{\,\,and}\\
\hat{\mathbf{n}}_{\beta}^{(j)}\cdot \hat{\mathbf{r}}_{ji} > \cos \theta, \\
\text{for any two patches $\alpha$ and $\beta$}
\end{cases}
 \\
0 
\end{cases}
\end{equation}
where $\hat{\mathbf{n}}_{\alpha}^{(i)}$ is a unit vector in the direction of patch $\alpha$ on particle $i$, and $\hat{\mathbf{r}}_{ij}= \mathbf{r}_{ij} / r_{ij}$. The angle $\theta = 7\degree$ controls the size of the patches. 

We simulate these particles using EDMD simulations \cite{hernandez2007discontinuous, smallenburg2013liquids}, where we again measure the pressure tensor (Eq.~\ref{eq:HS-pressure}). During these simulations, the temperature is kept fixed via an Andersen thermostat \cite{bookfrenkel}.

\section{Direct coexistence in the canonical ensemble}

\begin{figure}
    \includegraphics[width=0.45\textwidth]{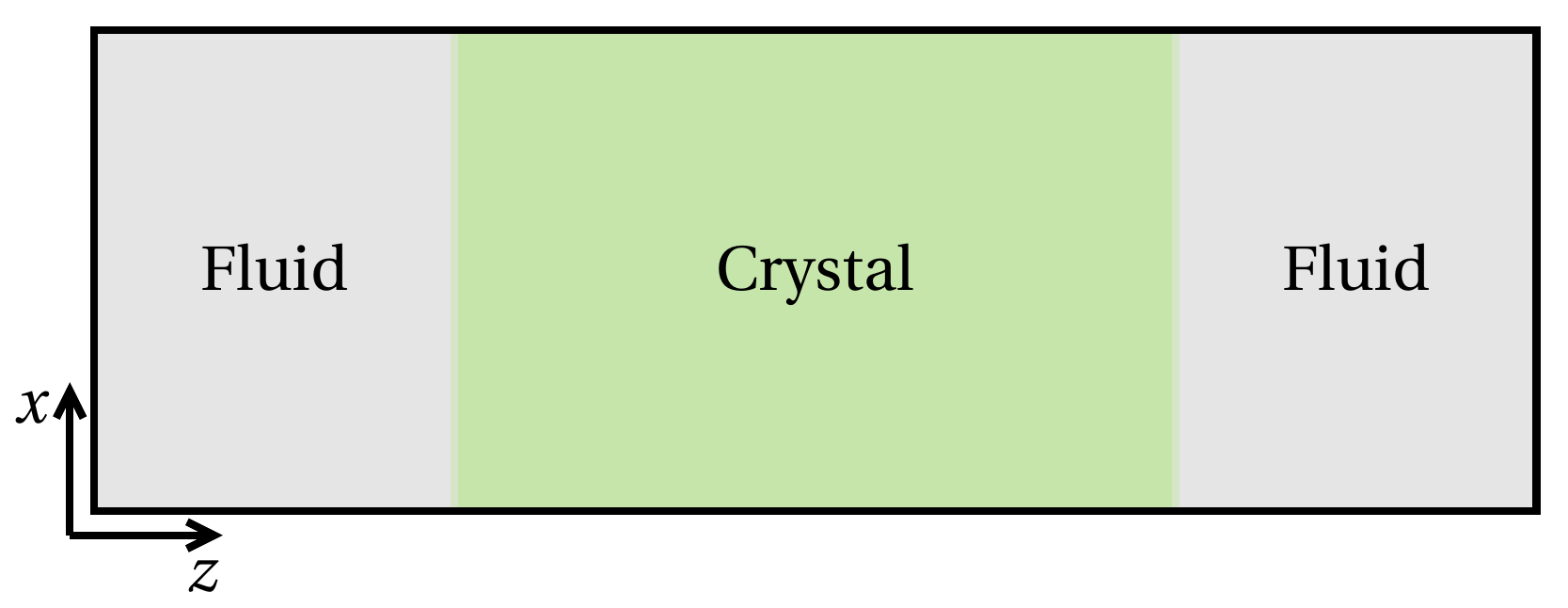}
    \caption{
    Sketch of the fluid-crystal coexistence in an elongated simulation box. Note that the simulation box is periodic in all directions.
    \label{fig:cartoon}
    }
\end{figure}

We consider a periodic simulation box elongated along the $z$-direction, containing a direct coexistence between a fluid and a crystal (see Fig. \ref{fig:cartoon}), in the $NVT$ ensemble. For simplicity, we first consider a monodisperse system for which the stable crystal phase has cubic symmetry (e.g. face-centered cubic). As a result, we can orient the crystal such that the $x$ and $y$ directions of our simulation are equivalent.
To minimize the overall interfacial area and hence the free energy, the most stable configuration of the interfaces will be such that they are oriented  perpendicular to the long $z$-axis of the box. Regardless of the simulation method used (e.g. Monte Carlo or molecular dynamics), standard methods exist to measure the overall pressure tensor $P_{ij}$ in the simulation box \cite{bookfrenkel}.

Let us assume that we have already measured the bulk equation of state of the crystal phase. In other words, for any given number density $\rho^X$ near the melting density $\rho^X_\mathrm{coex}$, we know the pressure of the undeformed crystal $\Pud(\rho^X)$. We can then take a crystal of any density (near where we expect the melting density to be),  and create a simulation box where it is in contact with a fluid as sketched in Fig. \ref{fig:cartoon}. In practice, creating this initial state can be done in a variety of ways. Depending on the chosen system size and the model under consideration, it may be sufficient to fill most of the box with the chosen crystal and leave some extra space on one side to facilitate melting. However, if the coexistence region is narrow, this may lead to a stretched crystal filling the entire simulation box. In that case, one strategy is to locally melt one half of the elongated box by e.g. raising the temperature or reducing the particle size, while keeping the particles in the other half of the simulation box fixed (either by pinning them in place or by greatly increasing their mass). For particles without hard-core interactions, it is also possible to place the particles in the fluid region randomly (followed by a rapid energy minimization to eliminate excessively strong interparticle forces). A final alternative is the separate equilibration of the fluid and crystal regions, followed by combining a fluid and a crystal configuration together into a system that contains an interface.

Regardless of how the initial configuration is created, the length of the long axis of the box should be chosen such that the overall density $\rho^\mathrm{global}$ lies within the coexistence region of the system under consideration. This can be checked by allowing the simulation to equilibrate using normal molecular dynamics or Monte Carlo schemes: if the global density is chosen too low, we expect the entire system to melt, while if it is too high, we expect it to freeze (assuming the crystal phase is denser than the fluid phase). In contrast, sufficiently deep within the coexistence region (and for sufficiently large systems), the equilibrium state should be a two-phase coexistence, with the amount of each phase determined by the lever rule. Some trial and error may be needed to find a global density that results in approximate half of the box being filled with crystal, in order to minimize finite-size effects that might result from thin slabs of either crystal or fluid.

In the geometry of Fig. \ref{fig:cartoon}, the periodic boundary conditions allow deformation of the crystal in only one direction: it can elongate or compress along the $z$ direction. The $x$ and $y$ directions are fixed by the periodic boundaries, which also prevent shear deformations 
\footnote{
In principle, one could imagine shearing the crystal phase in the $xz$ or $yz$ plane, by moving the interfacial crystal planes tangentially to the interface, without violating the periodic boundaries. However, this would induce a tangential stress in the crystal, which would need to be balanced by an opposite stress in the fluid phase to maintain mechanical equilibrium. Since the fluid phase cannot support tangential stresses, this cannot be a stable deformation in the applied geometry.
}. If our choice of $\rho^X_0$ results in an  unstrained crystal then the pressure tensor \textit{inside} the crystal phase is isotropic, i.e.
\begin{equation}
P^X_{ij} = \Pud(\rho^X) \delta_{ij},
\end{equation}
where $P^X_{ij}$ denotes the pressure of the crystal. In general, however, equilibration of the direct coexistence simulation will lead to a \textit{deformed} crystal, where the lattice spacing along the $z$-axis is stretched by a factor $\epsilon_{zz} = \rho^X_0 / \rho^X$, with $\rho^X_0$ and $\rho^X$ the initial and average values of the crystal density, respectively. The normal component of the pressure tensor $P^\mathrm{X}_{zz}$ inside the coexisting crystal phase can then be written as
\begin{equation}
P^\mathrm{X}_{zz}(\rho^X_0, \epsilon_{zz}) = P^\mathrm{ud}(\rho^X_0) - B_{zzzz}(\rho^X_0) \epsilon_{zz}  + \mathcal{O}(\epsilon_{zz}^2),
\end{equation}
where $B_{zzzz}$ is the effective elastic constant \cite{ray1989effective} of the crystal corresponding to a pure expansion along the $z$-axis. Mechanical equilibrium requires that the $\Pglobal_{zz}$ component of the pressure tensor is homogeneous throughout the system (see Appendix \ref{app:pressure}), and hence $\Pglobal_{zz} = P^\mathrm{X}_{zz}$. Hence, to determine the conditions where the crystal phase is undeformed ($\epsilon_{zz} = 0)$ we simply have to find the choice of $\rho_0^X$ where 
\begin{equation}
    \Pglobal_{zz}(\rho_0^X) = \Pud(\rho_0^X).
\end{equation}
In practice, this means that determining coexistence conditions requires that we find the crossing point between the functions $\Pglobal_{zz}(\rho_0^X)$, measured in direct coexistence simulations, and $\Pud(\rho_0^X)$, measured in the bulk crystal phase.

To see how this works, we will work through this method in detail for the hard-sphere model, and then show extensions to other model systems.

\section{Model 1: Hard spheres}

\subsection{Fluid-FCC crystal coexistence}

As a natural starting point for testing the direct coexistence method, we first focus on the fluid-FCC transition in monodisperse hard spheres.

\begin{figure*}
    \begin{tabular}{ccc}
    \includegraphics[width=0.375\textwidth,valign=m]{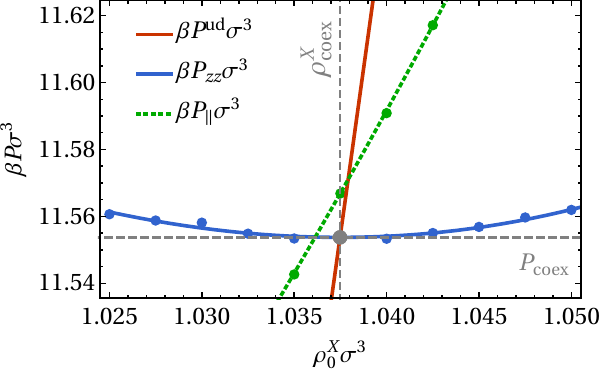}&&
    \includegraphics[width=0.575\textwidth,valign=m, raise=0.4cm]{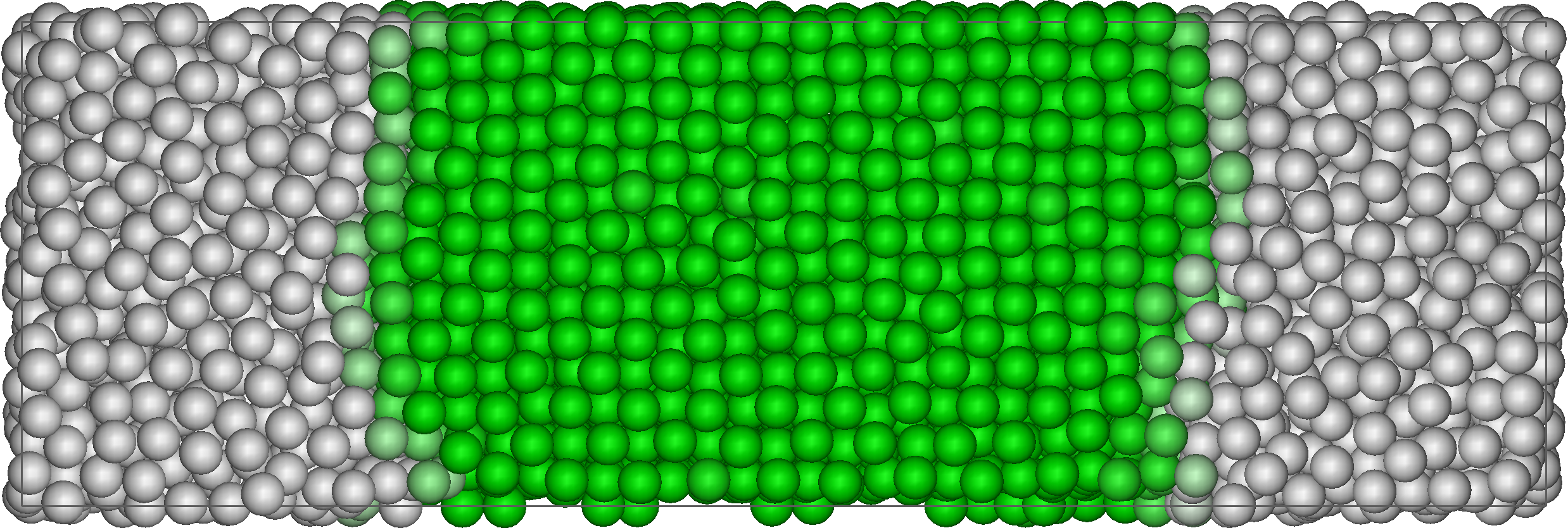} 
    \end{tabular}
    \caption{Direct coexistence approach for a hard-sphere system of $N=4116$ particles. The plot shows the behavior of the pressure $\Pglobal_{zz}$ normal to the interface as a function of the lattice spacing of the initial crystal $\rho^X_0$ (blue line). The coexistence point (gray dot) is determined as the crossing point of this line with the bulk equilibrium equation of state (red line). Note that at the point of equilibrium coexistence, the pressure component parallel to the interface ($P_\parallel$, green dashed line) is not the same as $\Pglobal_{zz}$, due to the stresses exerted by the interface. Statistical errors are on the order of the typical deviations of the points from the fitted lines. The snapshot shows a typical configuration from the direct coexistence simulation. As a guide to the eye, particles are colored based on the crystallinity of their local environment, using the averaged bond order parameter $\bar{q}_6$ \cite{lechner2008accurate}.
    \label{fig:HS}}
\end{figure*}

We first determine the bulk equation of state $\Pglobal_{zz}(\rho_0^X)$ in the vicinity of the melting point for a cubic FCC crystal of $N=1372$ particles. Next, we construct initial configurations for a range of densities $\rho^X_0 \sigma^3 \in \{1.025, 1.0275, 1.030, \ldots, 1.050\}$, by placing particles on an FCC lattice oriented with the square (100) face perpendicular to the interface in an elongated box chosen to be approximately three times longer in the $z$-direction than in the $x$ and $y$ directions. We then add additional empty space on one side of the crystal in the $z$-direction in order to reach an overall system density $\rho^\mathrm{global}\sigma^3 = 0.99$. In order to have similar finite-size effects in the bulk equation of state and the direct coexistence simulations, we use the same number of FCC unit cells along the shortest axis of the box in both simulations, resulting in $N=4116$ particles in the elongated box. 
After equilibration, the system reaches a stable fluid-crystal coexistence (see Fig. \ref{fig:HS}b for a typical snapshot). During the simulation, we measure the global stress tensor $\Pglobal_{ij}$. In Fig. \ref{fig:HS}a, we plot both $\Pglobal_{zz}(\rho^X_0)$ (blue line) and $\Pud(\rho^X_0)$ (red line) for a relatively small system size. The crossing point between these two lines then gives us the melting density $\rhocoex^X \sigma^3 = 1.03749$ and coexistence pressure $\beta \Pcoex \sigma^3 = 11.5524$ for this system size. 

Fig. \ref{fig:HS}a shows that the crossing point between $\Pglobal_{zz}$ and $\Pud$ essentially coincides with a minimum in $\Pglobal_{zz}$. This can be  understood when considering the fact that for each choice of $\rho^X_0$, the measured value of $\Pglobal_{zz}$ represents the pressure at which the fluid becomes metastable with respect to a crystal with this deformation. In equilibrium, the fluid will freeze as soon as there is \textit{any} crystal phase more stable than the fluid. Hence, the realization of the crystal that corresponds to the lowest coexistence pressure must correspond to the true equilibrium phase transition.

In principle, this provides another avenue for estimating the coexistence pressure. However, in practice, it is much harder to accurately determine the minimum in $\Pglobal_{zz}$ than its crossing point with $\Pud$. This is readily visible from Fig. \ref{fig:HS}a, as the steepness of the red line ($\Pud$) indicates that small errors in the measurement of $\Pglobal_{zz}$ will not strongly affect the predicted coexistence pressure.

\definecolor{FCCAcolor}{rgb}{0,0,0}
\definecolor{FCCBcolor}{rgb}{0.5,0,0.5}
\definecolor{HCPcolor}{rgb}{1,0.5,0}

\begin{figure}
    \begin{tabular}{lll}
     a) \hspace{0.3cm} \color{FCCAcolor}{\bf FCC$_\mathrm{A}$} & \hspace{0.6cm} \color{FCCBcolor}{\bf FCC$_\mathrm{B}$} & \hspace{0.8cm}\color{HCPcolor}{\bf HCP}\\
    \includegraphics[width=0.15\textwidth]{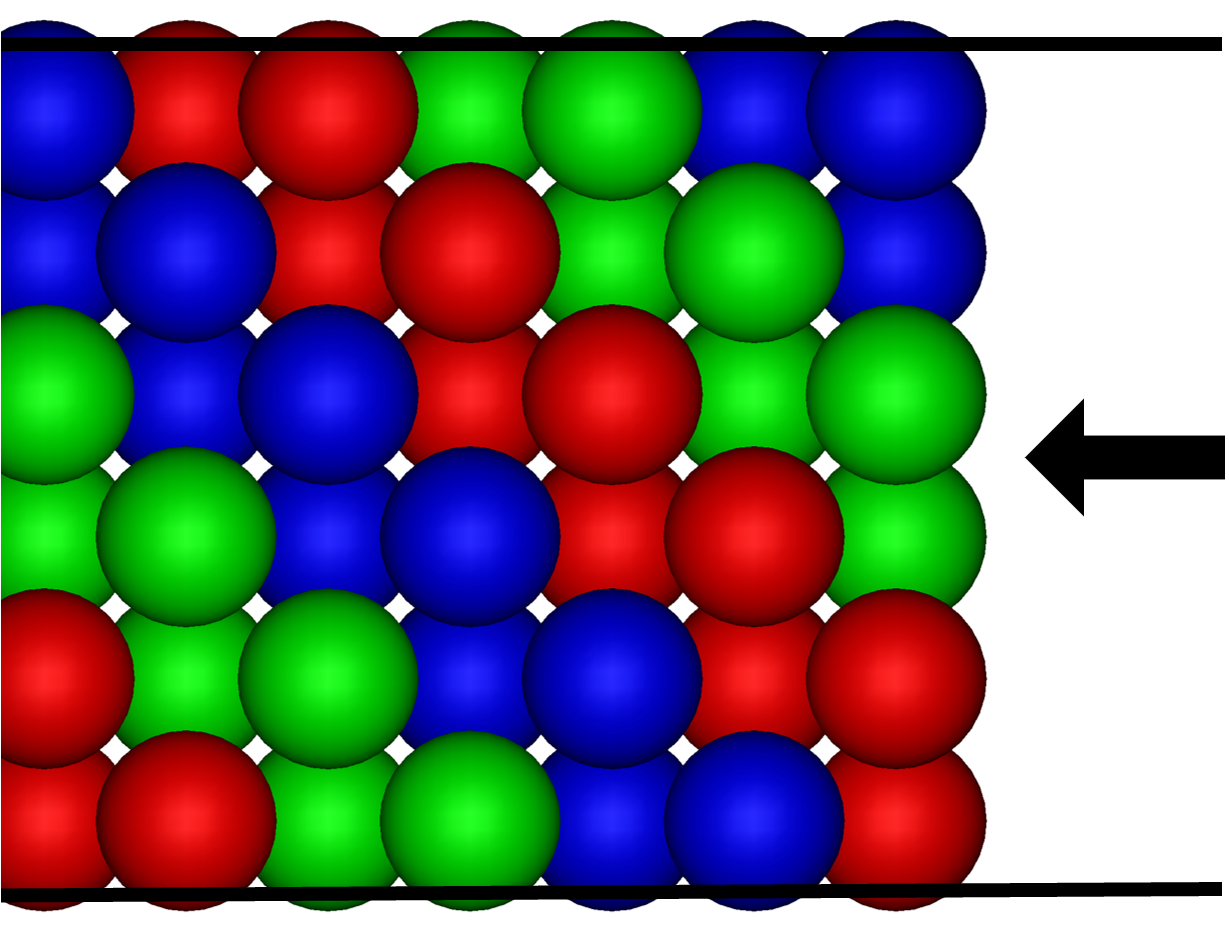} &
    \includegraphics[width=0.15\textwidth]{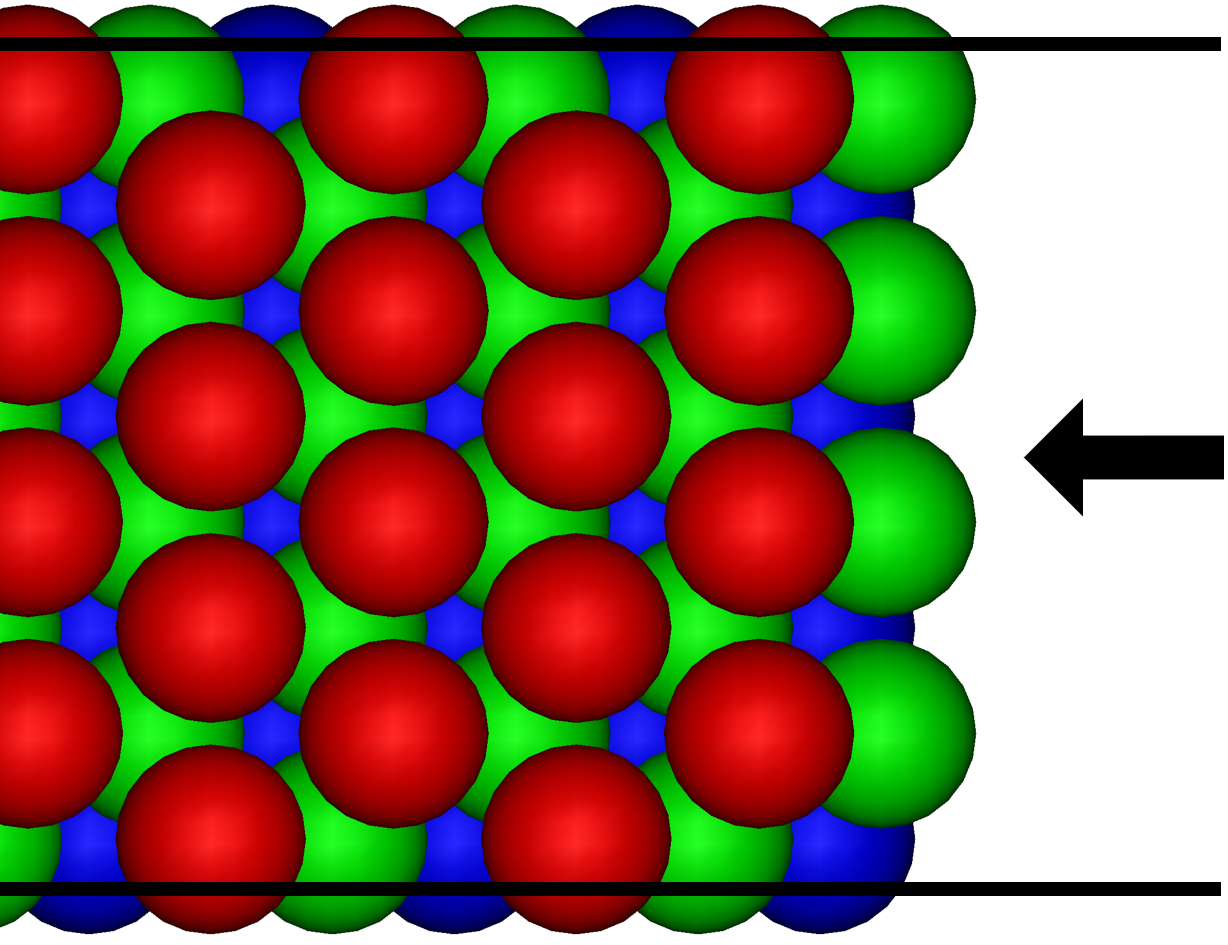} &
    \includegraphics[width=0.15\textwidth]{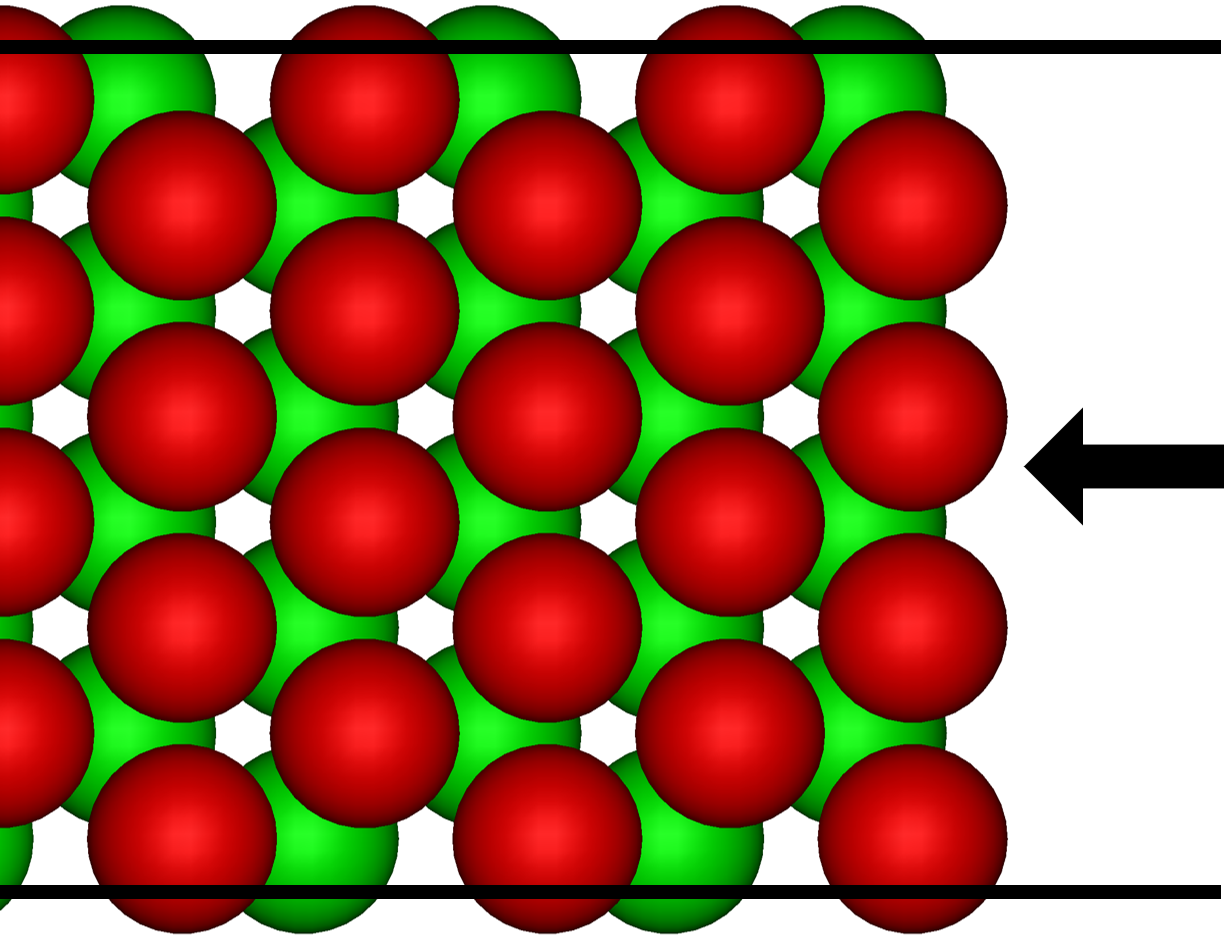} 
    \end{tabular}\\[0.5cm]
    \raggedright b)\hfill\\[-0.0cm]
    \includegraphics[width=0.45\textwidth,valign=t]{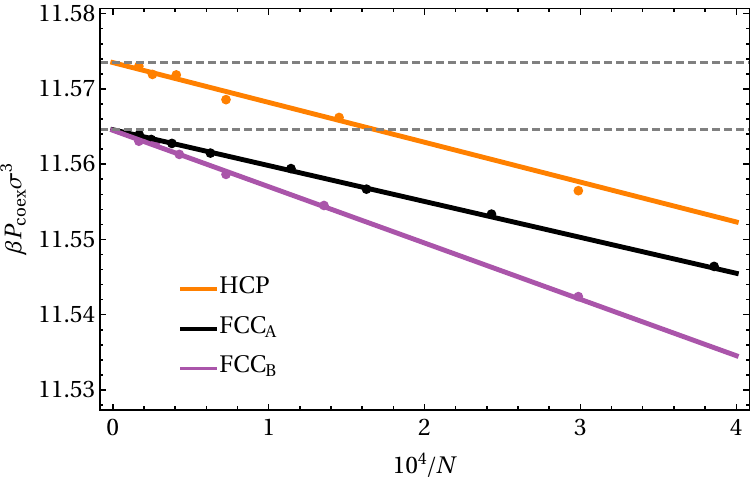}\\
    \raggedright c)\hfill\\[-0.0cm]
    \includegraphics[width=0.45\textwidth,valign=t]{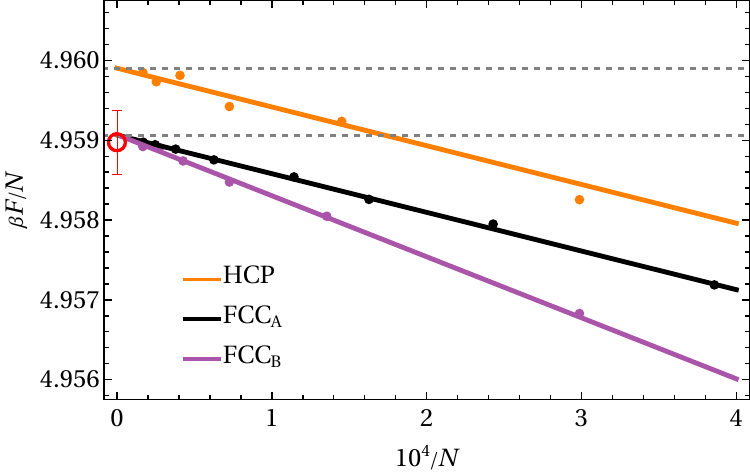} 
    \caption{\textbf{a)} Schematic images of the crystal orientations. In each image, the arrow points at the interface facing the fluid. \textbf{b)} Coexistence pressure as a function of  system size (characterized by the total number of particles $N$ in the direct coexistence simulations), for HCP and two different orientations of FCC. \textbf{c)} Helmholtz free energy of the crystal phase at density $\rho \sigma^3 = 1.0409$ (slightly above melting) as a function of system size. The red point represents the value obtained by \citet{bookfrenkel} and its corresponding error bar. For both a) and b), the solid lines are linear fits to the data for $N \geq 5000$, and the dashed lines indicate the values obtained by extrapolating these fits to $N\to\infty$.
    \label{fig:HS2}}
\end{figure}

Also shown in Fig. \ref{fig:HS}a is the behavior of $\Pglobal_\parallel = (\Pglobal_{xx}+\Pglobal_{yy})/2$ as measured from our direct coexistence simulations. We emphasize the deviation between $\Pglobal_{zz}$ and $\Pglobal_\parallel$ at the point of equilibrium coexistence. This difference can be directly linked to the free-energy cost associated with ``stretching" the interface, also known as the surface stress $f$ \cite{davidchack1998simulation}. Although not important to the determination of the coexistence conditions, this further shows why the assumption or requirement that the global pressure is isotropic in the direct coexistence simulation is not technically correct. We note, however, that in the limit of infinite system sizes, this deviation vanishes.

The coexistence pressure and melting density obtained from Fig. \ref{fig:HS}a contain finite-size effects. To quantify these effects and obtain a prediction for the infinite-system coexistence conditions, we repeat the same calculations for system sizes ranging from $N^\mathrm{global} \simeq 1500$ to $6\cdot 10^4$ particles \footnote{
Note that for these calculations, we also re-calculated the bulk equation of state for different crystal orientations and system sizes. However, in practice the finite size effects on the equation of state have a negligible effect on the overall determination of the coexistence conditions: repeating our calculations with the $Z_{S2}$ hard-sphere crystal equation of state by \citet{pieprzyk2019thermodynamic} yields essentially indistinguishable results, especially for larger system sizes.}. The simulations were run for simulation times of at least $10^5 \tau$, where $\tau = \sqrt{\beta m \sigma^2}$ is the time unit of our simulation. This is typically enough to obtain a good estimate of the coexistence conditions, especially for larger system sizes. However, some simulations for the smaller system sizes were run for up to 10 times longer to decrease noise.

We plot the resulting coexistence pressures in Fig. \ref{fig:HS2}a as a function of the inverse system size (black line). Extrapolating the behavior to infinite system size, we obtain $\beta\Pcoex\sigma^3 = 11.5645(5)$, which is in excellent agreement with the best known predictions in literature (see Table \ref{tab:compare}). Note that as expected, finite-size effects shift the observed coexistence pressure to lower values for smaller systems, as the periodic boundaries help stabilize the crystal phase.

\begin{table*}
\begin{ruledtabular}
\begin{tabular}{llllll}
\textbf{Source} & Method & $\rho_\mathrm{f}\sigma^3$ & $\rho_\mathrm{m}\sigma^3$ & $\beta P_\mathrm{coex}\sigma^3$  & $N$\\
\hline
  \citet{davidchack1998simulation}  & Direct coex. ($NVT$)  & 0.938       & 1.037     & 11.55(5)  & 10752 \\
  \citet{bookfrenkel}               & Free energy           & 0.9391      & 1.0376    & 11.567    & $\infty$\\
  \citet{fortini2006phase}               & Free energy           & 0.939(1)    & 1.037(1)  & 11.57(10) & -\\
  \citet{vega2007revisiting}        & Free energy           & 0.9387      & 1.0372    & 11.54(4)  & $\infty$ \\
  \citet{noya2008determination}     & Direct coex. ($NP_zT$)& 0.9375(14)  & 1.0369(33)& 11.54(4)  & 5184   \\
  \citet{zykova2010monte}           & Direct coex. ($NP_zT$)& 0.949       & 1.041     & 11.576(6) & $160000$\\    
  \citet{moir2021tethered}          & Free energy           & 0.93890(7)  & 1.03715(9)& 11.550(4) & $\infty$ \\
  \hline
  This work                         & Direct coex. ($NVT$)  & 0.93918(1)  & 1.03752(1) & 11.5645(5)& $\infty$
\end{tabular}
\end{ruledtabular}
\caption{Comparison of the predicted hard-sphere phase coexistence conditions to literature values. Note that all of these predictions neglect the effects of defects (see Discussion). In the last column, the dash(-) indicates that the treatment of system size was not reported.
\label{tab:compare}}
\end{table*}

In the above, we have made the choice to orient the FCC crystal with its square crystal plane facing the fluid. In principle, the coexistence conditions (in the thermodynamic limit) should be independent of the crystal orientation. To test this, we have repeated our calculation with the FCC crystal oriented such that the hexagonal planes in the crystal are aligned with the $xz$-plane of the box, as shown in Fig. \ref{fig:HS2}a. As a result, the plane facing the fluid is perpendicular to these hexagonal planes. The resulting coexistence pressures are shown as the purple line in Fig. \ref{fig:HS2}. As expected, for small systems the orientation matters, as the finite-size effects are different for different orientations of the crystal. However, in the limit of large systems, the two lines converge towards indistinguishable values. 

In principle, we could repeat the same calculation with the FCC crystal oriented such that the hexagonal plane faces the fluid. However, this orientation leads to an added complication: melting and reforming the surface allows for the introduction of stacking errors in the FCC structure, which results in a random hexagonally close-packed (rHCP) structure after sufficiently long simulations. Since our focus here is on the FCC crystal, we avoid this orientation.

\subsection{Fluid-HCP coexistence in hard spheres}
It is straightforward to extend our approach to crystals without cubic symmetry, for instance the hexagonal close packed (HCP) in hard spheres. For such non-cubic crystals, the lattice parameters of the stable crystal phase (i.e. the lengths and directions of the vectors spanning the unit cell) are generally dependent on the density. Hence, the determination of the equation of state should be done while taking into account the possibility of lattice deformations (e.g. in an isotension ensemble).  This then also provides the shape of the crystal lattice as a function of the density. The obtained crystal lattice for each density can then be directly used in the direct coexistence simulation, by adapting the shape of the simulation box in the $xy$ plane.  

To further test the sensitivity of our method, we explore the HCP-fluid coexistence in systems of hard spheres.  The HCP crystal in hard spheres is known to be metastable with respect to the FCC crystal, but is extremely close in free energy. Hence, its coexistence pressure with the fluid is expected to be slightly higher than that of the FCC phase.  As a first step to predicting this coexistence, we determine the pressure and lattice parameters of the HCP crystal as a function of density. Due to the hexagonal symmetry of the HCP lattice, the only parameter we have to determine is the ratio $c/a$ of the unit cell, where $a$ is lattice spacing inside the close-packed hexagonal layers and $c$ the height of the unit cell. For equilibrium hard-sphere crystals close to melting, this ratio is known to be close to the idealized value $\sqrt{8/3}$ \cite{pronk2003large}.

In order to continue using the same EDMD simulations in a constant-volume ensemble, we measure the lattice parameter $c/a$ by performing at each density simulations for several different values $\sqrt{8/3} - c/a \in \{0, 2.5\cdot 10^{-4}, ... 1.0\cdot10^{-3}\}$, and identifying the deformation for which the pressure tensor is isotropic (see Appendix \ref{app:pressure}). 

We then use the resulting lattice parameters as a function of density to initialize our direct coexistence simulations, where we orient the HCP crystal such that the hexagonal planes in the crystal are again aligned with the $xz$-plane of the box, as shown in Fig. \ref{fig:HS2}. We plot the resulting coexistence pressures in Fig. \ref{fig:HS2} along with the FCC results. As expected, we observe that the coexistence pressure for HCP is higher than that of FCC, by approximately $0.009 k_B T/\sigma^3$. 

\subsection{Calculating crystal free energies}

Direct coexistence simulations also provide a straightforward avenue to determine the free energy of crystal phases. At coexistence, the chemical potentials of the fluid and crystal phase coincide. Hence, knowing the chemical potential of the fluid also implies that we know the chemical potential of the crystal. The chemical potential of the fluid can be straightforwardly obtained from its equation of state via thermodynamic integration. To this end, we first determine the freezing density $\rhocoex^F$ from the coexistence pressure by using the mKLM hard-sphere fluid equation of state of Ref. \cite{pieprzyk2019thermodynamic}. Using the same equation of state, we then calculate the chemical potential via thermodynamic integration from an ideal gas \cite{bookfrenkel}:
\begin{eqnarray}
    \mu_\mathrm{coex} &=& \frac{F^F_\mathrm{coex}}{N} + \frac{P_\mathrm{coex}}{\rhocoex^F}\\
    \frac{\beta F^F_\mathrm{coex}}{N} &=& \log(\rhocoex^F \Lambda^3) -1 +
    \int_0^{\rhocoex^F} \mathrm{d} \rho^\prime
    \frac{\beta P(\rho^\prime) - \rho^\prime}
         {(\rho^\prime)^2}, \nonumber \\
\end{eqnarray}
with $\Lambda$ the thermal wavelength. Note that the value of $\Lambda$ does not affect the phase behavior, as it only results in a constant shift of the free energy in all phases. Hence, we choose to set it equal to $\sigma$ as is commonly done in free-energy calculations of hard spheres.

The Helmholtz free energy of the crystal at coexistence is then given by
\begin{equation}
    \frac{F^X(\rhocoex^X)}{N} = \mu_\mathrm{coex} - \frac{\Pcoex}{\rhocoex^X}.
\end{equation}
Using this reference value, we can calculate the free energy at any density inside the crystal regime via thermodynamic integration over the equation of state of the crystal:
\begin{equation}
    \frac{\beta F^X(\rho)}{N} =
    \frac{\beta F^X(\rhocoex^X)}{N} +
    \int_{\rhocoex^X}^{\rho} \mathrm{d}\rho^\prime 
    \frac{\beta \Pud(\rho^\prime)}{(\rho^\prime)^2}. \label{eq:eosintegration}
\end{equation}
Using this approach, we calculate the free energy of the crystal at a density of $\rho \sigma^3 = 1.0409$, where we can compare to the result of \citet{bookfrenkel} obtained using Einstein integration and finite-size scaling. We plot the results for both our FCC and HCP crystals in Fig. \ref{fig:HS2}b for different system sizes, and include the extrapolated infinite-size result of \citet{bookfrenkel} for FCC as a benchmark. Clearly, for both FCC orientations our free energies converge to the same free energy, while the HCP value is significantly higher. This allows us to calculate the free-energy difference between FCC and HCP, which we estimate to be $9.7\cdot 10^{-4} k_B T$ per particle at this density. This is in excellent agreement with past calculations using Einstein integration \cite{frenkel1984new, bolhuis1997entropy, mau1999stacking}, which estimate the difference to be approximately $0.001 k_B T$ per particle near melting.

\section{Model 2: Yukawa particles}

\newcommand\cincludegraphics[2][]{\raisebox{-0.3\height}{\includegraphics[#1]{#2}}}

\begin{figure*}
    \begin{tabular}{ccc}
    \includegraphics[width=0.375\textwidth,valign=m]{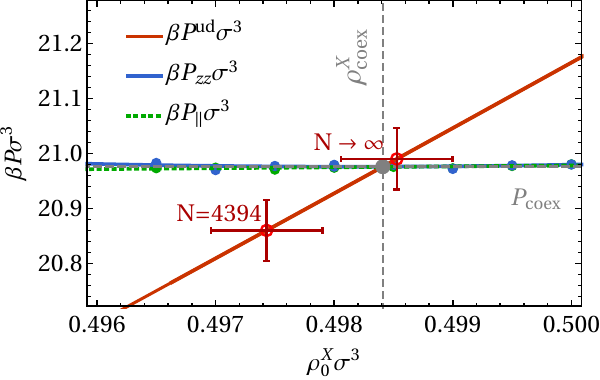}&&
    \includegraphics[width=0.575\textwidth,valign=m,raise=0.4cm]{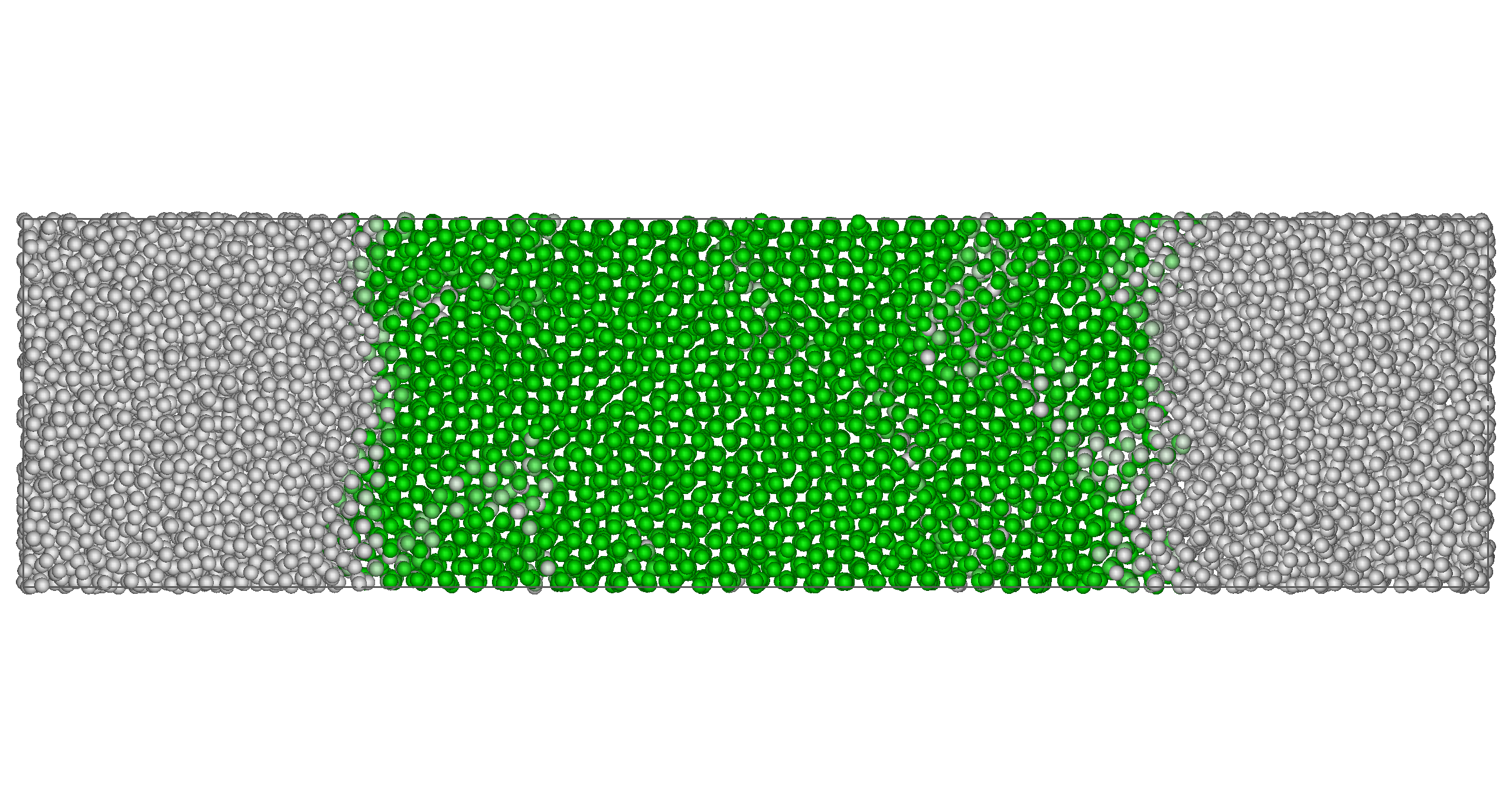} 
    \end{tabular}    \caption{Direct coexistence simulation of the Yukawa model with inverse screening length $\kappa \sigma = 4$, contact value $\beta \epsilon = 20$, and cutoff range $r_c/\sigma = 4.5$. The data in the plot is analogous to Fig. \ref{fig:HS}. The red circles indicates the predictions from free-energy calculations in the thermodynamic limit ($N\to \infty$) and for crystal system size $N=4394$. The latter corresponds to a crystal in a cubic box containing the same number of unit cells along each axis as used in the $x$ and $y$ directions of the long box simulations. As a guide to the eye, particles are colored based on the crystallinity of their local environment, using the averaged bond order parameter $\bar{q}_6$ \cite{lechner2008accurate}.
    \label{fig:yukawa}}
\end{figure*}

In order to illustrate the general nature of our methodology, we now turn our attention to a fluid-BCC coexistence of point Yukawa particles. We note that the fluid-BCC coexistence region in this model is expected to be very narrow \cite{hynninen2003phase}: the predicted width of the coexistence region is less than a percent of the melting density. Fluctuations in the amount of crystal phase in the direct coexistence simulation will therefore only weakly impact the densities of the two phases, and hence their free energies. As a result, we expect (and observe) larger fluctuations in the amount of crystal in this system in comparison to the hard-sphere system, necessitating long simulations to obtain good statistical averages. Similarly, large system sizes are required in order to avoid full crystallization or melting of the system as a result of these fluctuations. 

Our direct coexistence simulations are performed using systems of $N=17453$ particles, placed within a simulation box whose $z$-axis was approximately four times longer than the $x$ and $y$ axes. The total density $\rho^\mathrm{global} \sigma^3=0.4962$, which results in a coexistence where approximately half of the system is crystalline (see Fig. \ref{fig:yukawa}). In the initial configuration, half of the particles are placed on a BCC lattice with the $(100)$ crystallographic direction lying along the $z$-axis and 13 unit cells along the short sides. The other half are placed randomly in the remaining volume of the box. We perform a short energy minimization before the start of the run to reduce the initial forces between particles in the starting configuration. Additional configurations at different values of $\rho_0^X$ are then generated by (anisotropically) rescaling of the simulation box. The simulations were run for $2.5\cdot10^6\tau$.

The results of the direct coexistence approach are shown in Fig. \ref{fig:yukawa}, where we again determine the crossing between the $\Pglobal_{zz}$ and $\Pud$ as a function of the initial crystal density. 

To confirm our result, we also predict the phase coexistence using free-energy calculations (red circles in Fig. \ref{fig:yukawa}, see Appendix \ref{app:yukawa}), finding good agreement. Note that the narrow coexistence region also impacts the sensitivity of our free-energy-based predictions to statistical or systematic errors: a (reasonable) estimated error of $0.001 k_B T$ in the crystal free energy would give rise to a shift of $\Delta P \approx 0.05 k_BT/\sigma^3$ in the predicted coexistence pressure, giving rise to the large error bars in Fig. \ref{fig:yukawa}. This is approximately five times as large as the corresponding $\Delta P$ would be in the hard-sphere system. In other words, the narrow coexistence region makes it more cumbersome to obtain an accurate prediction for the coexistence conditions in both methodologies. Similarly, the coexistence pressure is rather sensitive to finite-size effects in the free-energy calculations. As shown in Fig. \ref{fig:yukawa}, the coexistence pressure shifts noticeably as we change the size of the crystal used in our free-energy calculations.

\section{Model 3: Patchy disks}

Finally, to demonstrate the applicability of this method to systems of anisotropic particles, we examine a two-dimensional model consisting of hard disks decorated with equally spaced attractive patches. In particular, we simulate systems of $N=4232$ particles, at global densities $\rho^\text{global}\sigma^2 \simeq 0.714$. We use a simulation box whose $z$-axis is approximately 2.5 times longer than the $x$-axis, initializing the system by adding extra empty space along the $z$-axis analogous to what was done for hard spheres. We run the simulations for a simulation time of $10^6\tau$. Measurements of the pressure tensor, and relative statistical errors, were obtained over 10 independent runs per point. 

\begin{figure*}
    \begin{tabular}{ccc}
    \includegraphics[width=0.375\textwidth,valign=t]{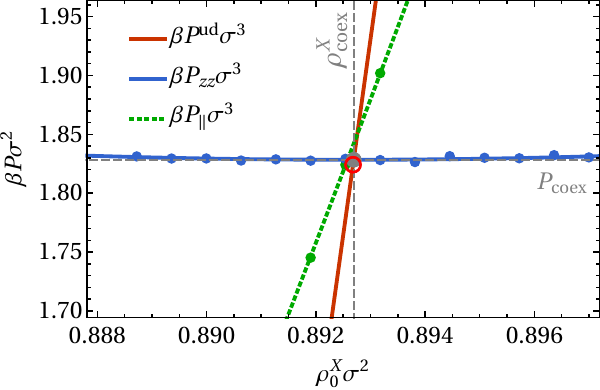}&&
    \includegraphics[width=0.5\textwidth,valign=t]{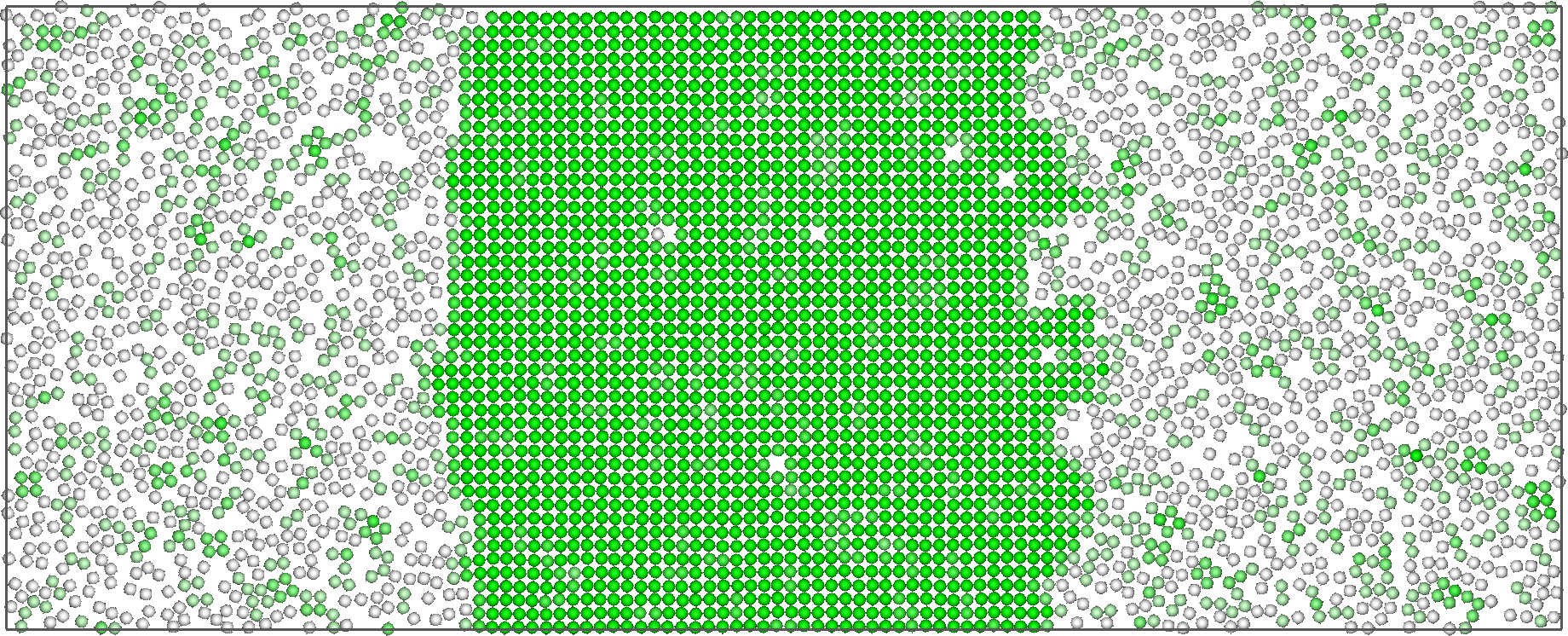} 
    \end{tabular}
    \caption{Direct coexistence simulation of the 4-patch Kern-Frenkel model at temperature $\epsilon/k_B T = -3.0$. The data in the plot is analogous to Fig. \ref{fig:HS}. The red circle indicates the result from free-energy calculations (statistical error bars are smaller than the point). The particles in the snapshot are colored based on the number of bonds formed by their patches.
    \label{fig:patchy}}
\end{figure*}

The direct coexistence results for this system are shown in Fig. \ref{fig:patchy}. The result is in close agreement with the prediction from the (significantly more cumbersome) free-energy calculations (red circle in Fig. \ref{fig:patchy}a, see Appendix \ref{app:patchy} for details). 

\section{Discussion and Conclusions}
We have presented a simple, accurate method to predict fluid-crystal coexistences based on direct coexistence simulations in the $NVT$ ensemble. As the algorithm is based on standard global pressure calculations, it can be used together with essentially any simulation method, and is hence compatible with any commonly used simulation package. 

As a brief recap, to find the fluid-crystal coexistence conditions for a monodisperse system, we:
\begin{enumerate}
    \item Determine the crystal equation of state $\Pud(\rho^X)$. This includes identifying the lattice parameters as a function of  density.
    \item Perform a series of direct coexistence simulations with different initial crystal densities $\rho^X_0$, and measure $\Pglobal_{zz}(\rho^X_0)$.
    \item Find the crossing point between $\Pglobal_{zz}(\rho^X_0)$ and $\Pud(\rho^X_0)$. The density and pressure of the crossing point are the melting density $\rhocoex^X$ and coexistence pressure $\Pcoex$ respectively.
    \item To obtain the freezing density, we can additionally measure the fluid equation of state $P^F(\rho)$, and find the density $\rhocoex^F$ at which the fluid pressure equals $\Pcoex$.
\end{enumerate}

This method avoids the stochastic nature of the $NP_{z}T$ approach of e.g. Refs. \cite{noya2008determination, espinosa2013fluid}, and therefore the need to run multiple simulations at the same state point to determine a melting probability. It is also significantly simpler than the interface pinning method \cite{pedersen2013computing}, which requires the introduction of a biasing potential and an order parameter to track the overall crystallinity of the system. Finally, in comparison to the approach of Davidchack and Laird \cite{davidchack1998simulation}, our method avoids the need to measure local stress profiles and manual adjustments of the simulation box to these measurements.

In comparison to free-energy calculations using e.g. the Frenkel-Ladd method \cite{frenkel1984new}, the direct-coexistence approach we propose here is much easier to implement. Most importantly, the direct coexistence approach allows for the determination of the coexistence densities and pressures without requiring a numerical integration over a series of simulation results. As such integrations can easily introduce numerical errors (due to a finite integration step size, the need to carefully choose integration limits, etc.) this immediately makes the direct coexistence approach significantly less error-prone. Additionally, free-energy calculations can present a number of pitfalls that can introduce errors in the result, which may be difficult to detect. For instance, in the Yukawa model studied here, simulations of the crystal close to melting allow for the spontaneous diffusion of particles within the lattice. If this occurs in the simulations associated with the Frenkel-Ladd integration (typically at low spring constants), special care must be taken to avoid a systematic error in the resulting free energy. Free-energy calculations also must explicitly take into account any configurational entropy associated with the crystal phase, as may occur in e.g. ice \cite{berg2007residual}, crystals of dumbbell-shaped particles \cite{marechal2008stability}, or quasicrystals \cite{fayen2024quasicrystal}. In contrast, this configurational entropy is inherently taken into account by the direct coexistence approach.

It is important to note that the direct coexistence method also comes with a few caveats. First, defects are not accurately taken into account in the methodology described above. In the direct coexistence simulations, point defects such as vacancies and interstitials are free to diffuse into and out of the crystal phase (as is visible in Fig. \ref{fig:patchy}), and hence for sufficiently long simulation times we would expect these simulations to correctly incorporate them. However, this may require long simulation times in practice. Moreover, we neglected the effects of defects on the bulk equation of state. In principle, this could be addressed with some additional effort, e.g. by measuring the defect concentration in the direct coexistence simulation (assuming it is large enough to be measurable), and checking the effect of these defects on the equation of state. We note, however, that taking into account defects in free-energy calculations also requires significant additional effort \cite{pronk2001point, de2021defects,van2020high} and is rarely done. 

Secondly, it should be noted that the direct coexistence approach is generally more computationally expensive than free-energy calculations. Equilibrating the explicit interface between the two coexisting phases and sampling its fluctuations over time requires simulations over longer time scales than sampling the behavior of the single-phase simulations required for a prediction of phase coexistence based on free energies. Moreover, the system sizes required to maintain a stable coexistence are significantly larger than those required to simulate a pure fluid or crystal in a reasonable approximation of the thermodynamic limit. This downside is partially addressed by the simplicity of the method, which means that the simulations can be performed by existing simulation codes that have already been well-optimized or adapted for parallel or GPU computing. However, if the model of interest has interactions that are computationally expensive, or requires very large system sizes to realize a stable interface, the computational cost may become prohibitive.

Finally, we point out that direct coexistence methods are not suitable for solid-solid transitions, as two unstrained crystals can typically not occupy the same simulation box \cite{bookfrenkel}. Nonetheless, in some cases, such as the case of hard-sphere HCP presented here, direct coexistence can still be useful if a metastable fluid-crystal coexistence can be simulated. The resulting crystal free energy at melting can then be used as a reference point for thermodynamic integration to other state points. However, for crystal phases that cannot form a metastable coexistence with a fluid, other methods would be required.

Despite these caveats, the $NVT$ direct coexistence method presented here is a highly accurate and convenient method for the prediction of fluid-crystal phase coexistences. As shown by our hard-sphere example, it is at least as accurate as free-energy calculations. Moreover, as we show with the Yukawa and patchy systems, the method is directly applicable to any fluid-crystal phase boundary. In short, for systems where a coexisting state can be equilibrated on reasonable time scales, $NVT$ direct coexistence is a powerful method that we expect to become a staple technique for the determination of crystal phase boundaries.

\section{Appendix}

\subsection{Pressure tensor in a coexisting system}
\label{app:pressure}
We consider a system of $N$ particles in a volume $V$ at temperature $T$, which exhibits a coexistence between a fluid and a (possibly strained) crystal. The box is elongated along the $z$-axis, and we will assume a slab-like coexistence geometry with interfaces perpendicular to $z$ (see Fig. 1 of the main paper).  We assume that the number of layers of crystal in the directions parallel to the interface is fixed. As the system is at constant volume and has periodic boundary conditions, the lattice parameters of the crystal in the directions parallel to the interface are constrained. 
To set the shape of the box in these directions, we choose it to be consistent with an equilibrium (strain-free) crystal at a density $\rho^X_0$. Note that during the direct coexistence simulation, the crystal will not necessarily remain unstrained as the  lattice constant in the $z$-direction can change.  However,  assuming that the crystal does not undergo any major rearrangements, the lattice constants in the $x$ and $y$ are fixed by the choice of $\rho^X_0$. Additionally, the surface area $A(\rho^X_0)$ of a single interface is trivially determined by the box size in the directions perpendicular to $z$.

We denote the number of particles in the fluid and crystal phase as $N^F$ and $N^X$, respectively, and use the same superscripts for their respective volumes $V^F,V^X$, number densities $\rho^F, \rho^X$, etc. 
Following standard conventions in dealing with systems with interfaces, we assume the interface to be a flat dividing surface perpendicular to the $z$-axis, with zero volume but potentially a non-zero number of particles $N^S$ associated with it, such that:
\begin{eqnarray}
N &=& N^F+N^X+N^S \\
V &=& V^F+V^X.
\end{eqnarray}
Without loss of generality, we choose the equimolar surface as our dividing surface, which is characterized by $N^S = 0$.

We can write down the total Helmholtz free energy of the system as
\begin{eqnarray}
    F^\mathrm{total}(N,V,  \rho^X_0; N^X, V^X) = 
    F^F(N^F, V^F)  \nonumber \\
    + F^X(N^X, V^X, \rho^X_0) + 
    2 \gamma(\mu, \rho^X_0) A(\rho^X_0).\label{eq:Ftot}
\end{eqnarray}
Here the semicolon in the functional dependence of $F^\mathrm{total}$ separates the variables that are externally fixed ($N,V, \rho^X_0$) and the variables that are chosen by the system itself ($N^X, V^X$) based on a minimization of its free energy. Additionally,  $\gamma$ is the interfacial free energy, which is generally dependent on both the chemical potential and the lattice spacing of the crystal in the directions parallel to the interface (denoted by its dependence on $\rho^X_0$), and the factor 2 arises due to the presence of two interfaces. Note, however, that our choice of the equimolar dividing surface imposes that 
\begin{equation}
    \diffix{\gamma}{\mu}{\rho_0^X} = N^S/A = 0.
\end{equation}

Minimizing the free energy with respect to $N_X$ yields
\begin{equation}
    0 = \diffix{F^\mathrm{total}}{N^X}{N,V,V^X,\rho^X_0} = 
    -\mu^F(\rho^F) + \mu^X(\rho^X_0; \rho^X),
\end{equation}
where $\mu^F$ and $\mu^X$ denote the chemical potentials of the two phases. We can rewrite this as
\begin{equation}
    \mu^F(\rho^F) = \mu^X(\rho^X_0; \rho^X), 
\end{equation}
confirming chemical equilibrium between the coexisting phases.

Similarly, minimizing the free energy with respect to $V_X$ yields
\begin{equation}
    0 = \diffix{F^\mathrm{total}}{V^X}{N,V,N^X,\rho^X_0} = 
    P^F(\rho^F) - P_{zz}^X(\rho^X_0; \rho^X),
\end{equation}
with $P^F$ the (isotropic) pressure of the fluid, and  $P_{zz}^X$ the pressures of the crystal phase along the $z$-direction. Hence, we also find mechanical equilibrium along the $z$-axis:
\begin{equation}
    P^F(\rho^F) = P^X_{zz}(\rho^X_0; \rho^X) \equiv \Pglobal_{zz}(\rho^X_0).  \label{eq:mechanicalequilibrium}
\end{equation}

The pressure of the crystal along the $z$-axis can be written more explicitly by taking into account the deformation of the crystal away from its equilibrium shape at density $\rho^X_0$. In the direct coexistence simulation, the crystal lattice can deform in response to any pressure imbalance between the fluid and the crystal. In particular, the crystal can either expand or compress along its $z$-axis, changing its density $\rho^X$ away from $\rho^X_0$. Note that since the fluid can only exert a net force along the $z$-axis of the box, it cannot induce an overall shear of the crystal parallel to the interface. Hence, we only have to consider deformations of the crystal phase characterized by a uniaxial strain $\epsilon_{zz}$, which (up to linear order in the deformation) can be written as
\begin{equation}
    \epsilon_{zz} = 1-\frac{\rho^X}{\rho^X_0}.
\end{equation}
The pressure of the crystal is therefore given by
\begin{eqnarray}
    P_{zz}^X(\rho^X,\rho^X_0) &=& \Pud(\rho^X_0) + \partdev{P_{zz}}{\epsilon_{zz}} \epsilon_{zz} + \mathcal{O}(\bm{\epsilon}^2) \\
    &=& \Pud(\rho^X_0) - B_{zzzz} \epsilon_{zz} + \mathcal{O}(\bm{\epsilon}^2), \label{eq:Pzz}
\end{eqnarray}
where $\Pud$ is the (isotropic) pressure of the undeformed equilibrium crystal, and $B_{zzzz}$ is the elastic constant associated with uniaxial compression or expansion along the $z$-axis, defined as
\begin{equation}
B_{zzzz} = \partdev{\sigma_{zz}}{\epsilon_{zz}},
\end{equation}
where $\bm{\sigma} = -\mathbf{P}$ is the stress tensor. Note that $B_{zzzz}$ depends on the orientation of the crystal, as different orientations of the crystal will cause the deformation to occur along different crystal directions.

Equation \ref{eq:Pzz} immediately implies that $P^X_{zz} = \Pud$ if and only if the crystal is unstrained. Combined with Eq. \ref{eq:mechanicalequilibrium}, this demonstrates that we can find the coexistence point between an unstrained crystal and the fluid by finding the density $\rho^X_0$ such that $\Pglobal_{zz}= \Pud$.

\subsection{Lattice parameters of the hard-sphere HCP crystal}
\label{app:hcp}

For the hexagonally close-packed (HCP) crystal of hard spheres, the shape of the lattice is dependent on density \cite{pronk2003large}. We denote the spacing between two neighboring particles inside a hexagonal layers as $a$, and the height of the HCP unit cell perpendicular to the hexagonal layers $c$ (such that $c$ is twice the spacing between two hexagonal layers). Then at close packing, the ratio $c/a = \sqrt{8/3}$. At lower densities, this ratio deviates slightly from the close-packing value.

\begin{figure}
\begin{tabular}{lll}
 a) & \\ &
    \includegraphics[width=0.9 \linewidth]{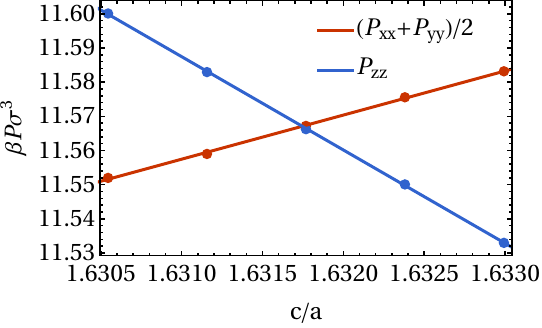} \\
b) & \\ & 
\includegraphics[width=0.9 \linewidth]{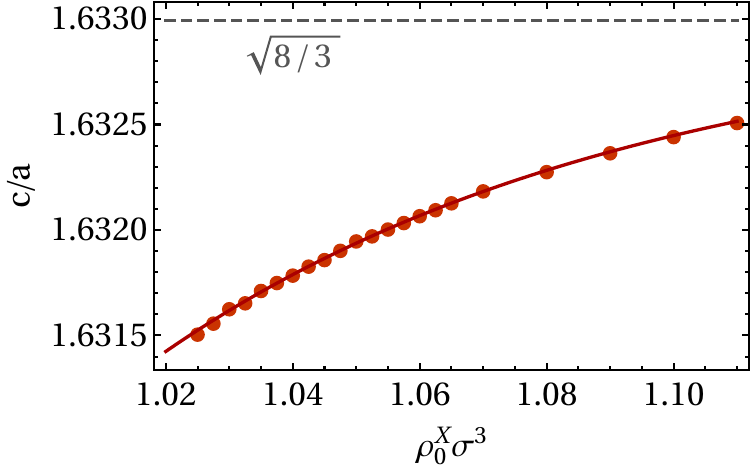}     
\end{tabular}
    \caption{a) Pressure tensor components along different box axes for an HCP crystal of $N=20160$ hard spheres at number density $\rho \sigma^3 = 1.0375$ with varying values of the lattice parameter $c/a$. The equilibrium value of $c/a$ is determined as the point where the two lines cross. b) Behavior of $c/a$ as a function of density at $N=20160$. The solid line is the fit from Eq. \ref{eq:hcpdef}. The dashed gray line indicates the close-packing value $c/a = \sqrt{8/3}$.
    \label{fig:hcpdeformation}
    }
\end{figure}

To measure this deviation, we perform event-driven molecular dynamics simulations of the HCP crystal for a range of densities close to coexistence, and for a small range of ratios $c/a =  \sqrt{8/3} - \delta$, with $\delta \in [0, 10^{-3}]$. During the simulation, we measure the pressure tensor. The equilibrium value of $c/a$ for a given density is then determined as the point where the pressure tensor becomes isotropic. A typical example is shown in Fig. \ref{fig:hcpdeformation}a. The direct coexistence method simulations are then initialized with the appropriate lattice shape associated with the initial crystal density. We find that near coexistence, the value of $c/a$ is well-approximated by
\begin{equation}
    \frac{c}{a} = \sqrt{\frac{8}{3}}\left(1 - 676 \exp(-13.2 x)\right), \label{eq:hcpdef}
\end{equation}
as shown in Fig. \ref{fig:hcpdeformation}b.

\subsection{Free-energy calculations for Yukawa particles}
\label{app:yukawa}
In order to verify the coexistence values obtained from the direct coexistence simulations in the Yukawa system, we additionally determine the coexistence conditions via a free-energy route.
To this end, we first determine the equation of state of the bulk fluid and crystal phases using Monte Carlo (MC) simulations in the $NVT$ ensemble. These simulations are all initialized as a perfect BCC crystal of 2000 particles. Additionally, we use MC simulations in the $NPT$ ensemble of a fluid of 2000 particles to determine the equation of state of the metastable fluid.  

To determine the free energy of the fluid phase, we use thermodynamic integration of the equation of state to obtain the free energies as a function of density, using the ideal gas as a reference system \cite{bookfrenkel}:
\begin{equation}
	\frac{\beta F}{N} = \log(\rho\Lambda^3)-1+ \int_{0}^{\rho}\frac{\beta P(\rho^\prime) - \rho^\prime}{{\rho^\prime}^2}\mathrm{d}\rho^\prime. \label{eq:fluidintegration}
\end{equation}
Here, $\Lambda$ is the thermal De Broglie wavelength, which we again set equal to $\sigma$. To perform the integral, we fit the equation of state of the fluid using the virial expansion up to 10th order, for which we calculated $B_2$ analytically.

For the crystal, we again use thermodynamic integration (analogous to Eq. \ref{eq:eosintegration}), using a 6th order polynomial fit to the equation of state and starting from a reference free energy at effective packing fraction $\eta=\pi \rho\sigma^3 / 6=0.29$. We obtain this reference free energy using Einstein integration \cite{frenkel1984new} and correct for finite-size effects by considering systems of 686, 1024, 1458, 2000, 2662, 3456, and 4394 particles \cite{polson2000finite}. 
In this approach, the absolute free energy of the crystal is determined as a thermodynamic integration between a reference system and the crystal of interest. The reference system consists of an Einstein crystal of non-interacting particles, which are tied to their lattice sites via harmonic springs with a spring constant $\alpha$. To this end, we perform a series of MC simulations with an effective Hamiltonian given by
\begin{equation}
H(\lambda) = (1-\lambda)U_\mathrm{Yuk}(\mathbf{r}^N) + \lambda U_\mathrm{Ein}^r(\mathbf{r}^N),    \label{eq:yukhamiltonian}
\end{equation}
where $\lambda$ is a parameter that tunes between the Yukawa crystal ($\lambda = 0$) and the Einstein crystal ($\lambda = 1$), and $U_\mathrm{Yuk}$ is the total interaction energy of the system resulting from the Yukawa pair interactions. $U_\mathrm{Ein}^r$ is the energy resulting from the springs binding the particles to their lattice sites, given by
\begin{equation}
    U^r_\text{Ein} = \frac{\alpha}{\sigma^2}\sum_i(\mathbf{r}_i-\mathbf{R}^{(i)}_0)^2, \label{eq:UEinR}
\end{equation}
where $\mathbf{R}^{(i)}_0$ are the equilibrium positions in the ideal lattice.
During the simulations, the center of mass is kept fixed \cite{bookfrenkel}.

The free energy of the interacting Yukawa crystal is then determined as \cite{bookfrenkel}
\begin{eqnarray}
\frac{\beta F}{N} = 3\log \frac{\Lambda}{d_\alpha} + \frac{1}{N} \log \frac{\rho d_{\alpha}^3}{N^{3/2}} - \frac{\beta}{N}\int_0^1 \mathrm{d}{\lambda}\left\langle \frac{\partial H}{\partial \lambda}\right\rangle_\lambda,
\end{eqnarray}
with $d_\alpha = \sqrt{\pi \sigma^2/\beta \alpha}$ the typical displacement of a particle in the Einstein crystal. Here, the first term represents the free energy of the Einstein crystal, the second term incorporates corrections due to the fixing of the center of mass \cite{bookfrenkel}, and the integral term represents the free-energy difference between the Einstein and Yukawa crystals (with fixed centers of mass). The subscript $\lambda$ in the integrand indicates that the measurement of $\frac{\partial H}{\partial \lambda}$ is done in a simulation where the parameter in Eq. \ref{eq:yukhamiltonian} is set to $\lambda$.

We use a spring constant of $\beta\alpha=34$ for the Einstein crystal, and, for each system size, perform the numerical integration using a 10-point Gauss-Legendre quadrature \cite{bookfrenkel} and estimate the error using an additional 11 points from the Gauss-Kronrod rule.

Using the fluid and crystal free energies, we finally find the equilibrium coexistence point by determining the conditions where the two phases have equal pressures and chemical potentials via a common-tangent construction. Error bars are estimated by varying the chosen integration paths (e.g. changing the reference density of the crystal, obtaining the fluid free energy by integrating over interaction strength rather than density, and varying the maximum spring constant), and examining the variation in the resulting free energies.

\subsection{Free-energy calculations for patchy particles}
\label{app:patchy}

For the patchy particles, we again confirm our direct coexistence results by predicting the phase transition via a free-energy route. For the equation of state of the fluid we perform EDMD simulations of $N=2116$ particles for a time of $2\cdot10^5\tau$ after equilibrating the system for $2\cdot10^4\tau$. At low densities we perform longer simulations, to ensure sufficient statistics.
Pressure values at each state point are averaged over 10 independent runs, and statistical error is also estimated. The fluid free energy is again calculated using Eq. \ref{eq:fluidintegration}, using a weighted fit on the integrand function using a 19-th order polynomial on 75 points, constraining the constant term to the analytically known second virial coefficient 
\begin{equation}
    B_2 = \pi\sigma^2/2 \left\{1 - \left(\exp{\beta\epsilon}-1\right) n_\text{p}^2 \frac{\delta^2}{\pi^2} \left[\left(\frac{\lambda_\text{p}}{\sigma}\right)^2-1\right]\right\},
\end{equation}
adapted from Ref. \onlinecite{dorsaz2012spiers} to the case of two-dimensional particles.

For the square crystal phase we calculate the free energy at packing fractions $\eta=0.70, 0.72,0.73$ using Einstein integration\cite{frenkel1984new}, using Monte Carlo simulations. For these anisotropic particles, in the Einstein crystal both the position and the orientation of each particles are tied to a reference point. In addition to the positional springs of Eq. \ref{eq:UEinR}, we now additionally include a constraining potential for the orientations:
\begin{eqnarray}
    U^\theta_\text{Ein} &=& \alpha\sum_i \sin^2\left(\frac{n_\text{p}(\theta_i-\theta^{(i)}_0)}{2}\right)
\end{eqnarray}
where $\theta_i$ is the orientation of particle $i$, and $\theta^{(i)}_0$ is its  current orientation in the ideal lattice. We then perform a series of simulations with $\alpha$ varying from $0$ to $\alpha_\mathrm{max} = 10^4$ and measure the mean values of both of the above expressions during each simulation, in a system interacting through the total potential $U_\text{KF}+U^r_\text{Ein}+U^\theta_\text{Ein}$. The free energy of the patchy square crystal (with $n_p = 4$) is then given by:
\begin{eqnarray}
\frac{\beta F}{N} &=& \log \frac{(\beta \alpha_\mathrm{max})^{3/2}\Lambda^2}{\sigma^2\sqrt{\pi}} + 
\frac{1}{N} \log\frac{\pi \rho \sigma^2}{N \beta \alpha_\mathrm{max}} \nonumber\\
&+& 2 \beta \epsilon - \frac{\beta}{N}\int_{0}^{\alpha_\text{max}}d\alpha\left\langle \frac{U^r_\text{Ein}+U^\theta_\text{Ein}}{\alpha} \right\rangle_\alpha,
\end{eqnarray}
where we have assumed that $\alpha_\mathrm{max}$ is large enough to ensure that when $\alpha = \alpha_\mathrm{max}$ all particles remain bonded to their four neighbors throughout the simulation, and the deviations of particles from their lattice sites are small enough that $V^\theta_\mathrm{Ein}$ is effectively harmonic.

We perform the integration from the Einstein crystal by using a 50-point Gauss-Legendre quadrature, estimating and propagating the statistical error over 10 independent runs per each point. We performed Monte Carlo (MC) NVT simulations of $N=2116$ particles for $10^6$ cycles, with a constrained center of mass.
Finally, analogously to the Yukawa system, we obtain the free energy as a function of the density by integrating along the equation of state (Eq. \ref{eq:eosintegration}), starting from the point at $\eta=0.70$. We then obtain the coexistence conditions via a common tangent construction, using both the fluid and crystal free energies. The error is estimated by considering the statistical error on the free energies and the numerical error on its derivative.

We did not perform finite-size analysis for the patchy system.

\section*{Supplementary Material}
The Supplementary Material contains a sample LAMMPS script for direct coexistence simulations of the Yukawa system.

\section*{Data availability statement}
The authors declare that the data supporting the findings of this study are available within the article and supplemental material, as well as a data package published on Zenodo: \href{http://doi.org/10.5281/zenodo.11259762}{http://doi.org/10.5281/zenodo.11259762}.

\section*{Acknowledgements}
We thank Alfons van Blaaderen, Giuseppe Foffi, and Rebecca Smaal for fruitful discussions. The authors acknowledge the use of the Ceres high-performance computer cluster at the Laboratoire de Physique des Solides to carry out the research reported in this article. L.F., M.d.J., and G.D. acknowledge funding from the Vidi research program with project number VI.VIDI.192.102 which is financed by the Dutch Research Council (NWO). F.S. acknowledges funding from the  Agence Nationale de la Recherche (ANR), grant ANR-21-CE30-0051. 

\bibliography{refs}

\end{document}